\documentclass[a4paper,11pt]{article}
\pdfoutput=1 

\usepackage{jinstpub} 

\usepackage{lineno}

\title{Using Convolutional Neural Networks to Reconstruct Energy of GeV Scale IceCube Neutrinos}

\author{Jessie Micallef}
\affiliation{Michigan State University,\\567 Wilson Rd, Lansing, MI 48824, USA}

\emailAdd{micall12@msu.edu}

\abstract{The IceCube Neutrino Observatory, located under 1.4 km of Antarctic ice, instruments a cubic kilometer of ice with 5,160 optical modules that detect Cherenkov radiation originating from neutrino interactions. The more densely instrumented center, DeepCore, aims to detect atmospheric neutrinos at 10-GeV scales to improve important measurements of fundamental neutrino properties such as the oscillation parameters and to search for non-standard interactions. Sensitivity to oscillation parameters, dependent on the distance traveled over the neutrino energy (L/E), is limited in IceCube by the resolution of the arrival angle (which determines L) and energy (E). Event reconstruction improvements can therefore directly lead to advancements in oscillation results. This work uses a Convolutional Neural Network (CNN) to reconstruct the energy of 10-GeV scale neutrino events in IceCube, providing results with competitive resolutions and faster runtimes than previous likelihood-based methods.}

\keywords{Analysis and statistical methods; Cherenkov detectors; Neutrino detectors}

\collaboration[c]{on behalf of the IceCube Collaboration$^*$\note[*]{Full author list and acknowledgments are available at \href{https://icecube.wisc.edu/collaboration/authors/\#collab=IceCube&date=2021-05-18&formatting=web&tag=VLVnT+2021}{icecube.wisc.edu}.}}

\proceeding{9$^{\text{th}}$ Very Large Volume Neutrino Telescope Workshop (VLVnT-2021)\\
  18-21 May 2021\\
  Valencia, Spain}

\notoc
\begin{document}
\maketitle
\flushbottom

\section{Introduction}
\label{sec:intro}

The IceCube Neutrino Observatory, located at the South Pole, detects atmospheric neutrinos that have interacted in the ice and bedrock. IceCube uses 5160 Digital Optical Modules (DOMs), attached to cables called strings that are arranged in a hexagonal array, to detect the Cherenkov light from neutrino interactions \cite{Aartsen:2016nxy}. The center-most DOMs are more densely arranged, in an array called DeepCore, to extend the detection energy down to GeV scales \cite{DeepCore}. These 10-GeV scale neutrinos are at energies higher than accelerator and reactor experiments can reach, giving a unique look at the neutrino oscillation region at these energies. Likelihood-based reconstruction methods are currently used to reconstruct the energy (E) and arrival angle (determines L), which are the necessary variables to constrain the oscillations parameters. An alternate solution is to apply machine learning methods to improve the resolution and runtime of the energy reconstruction.

\section{Convolutional Neural Network Architecture and Training}
\label{sec:CNN}

Neutrinos can be detected anywhere within the IceCube detector volume, so that events are spatially invariant, which is a symmetry that convolution neural networks  (CNNs) excel at identifying \cite{Krizhevsky:cnn, kaudererabrams2017quantifying}. IceCube has successfully implemented a CNN for high-energy events (> 100 GeV) \cite{DNN_Mirco, Huennefeld:2017tT}, but a CNN must be specifically optimized to handle low-energy neutrinos because the sparse DOM layout and the limited scattered light at these energies provide less input information for reconstruction.

\begin{figure}[htbp]
\centering 
\includegraphics[scale=0.25]{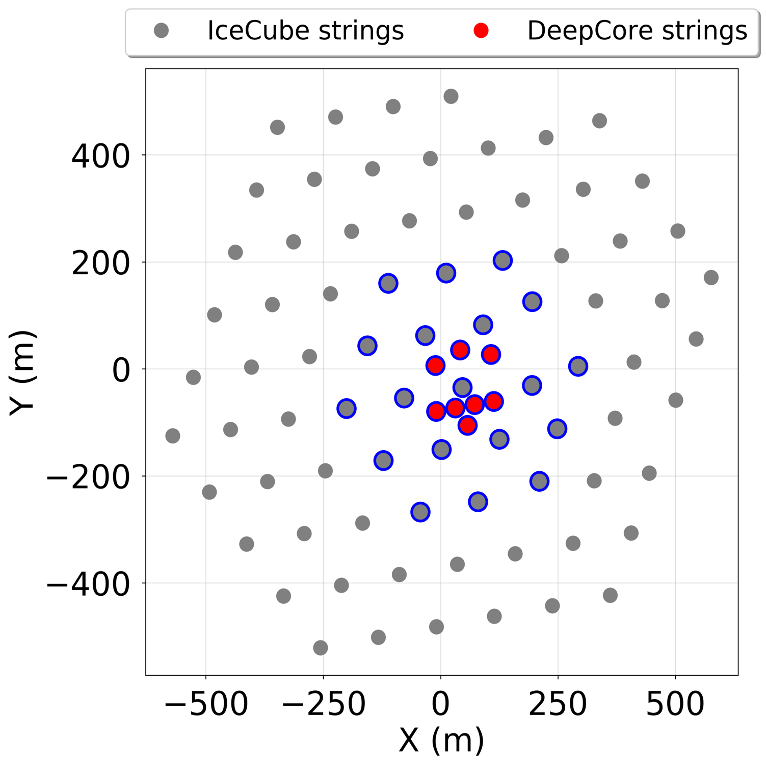}
\caption{\label{fig:TopView} Top view of the IceCube detector, with DeepCore strings in red, IceCube strings in gray, and a blue highlight on all strings used for the low-energy CNN.}
\end{figure}

The low-energy CNN only uses the 8 DeepCore strings and nearby 19 IceCube strings, focusing the CNN on the low-energy optimized region of the detector (Figure \ref{fig:TopView}). All hits per DOM are summarized into 5 charge and timing variables, which are separately given to the CNN for the DeepCore and IceCube strings because they have different z-spacing between DOMs, creating a two-branched network architecture. The CNN then uses its convolutional kernel to span the vertical, or z-depth, of the string; no xy-plane convolution is applied since the DeepCore strings are in an irregular geometry. After 8 convolutional layers, each followed by a dropout and batch normalization layer (see Ref. \cite{Jessie-proceedings} for more detail), the two CNN branches are concatenated to train on one fully connected layer, giving a single output: the reconstructed energy.

To train the CNN, a sample of more than 6 million unweighted, charge-current (CC) $\nu_\mu$ Monte Carlo (MC) events was generated in GENIE \cite{GENIE}, a neutrino MC generator with an emphasis on few-GeV neutrino interaction physics. Since $\nu_e$ events look like $\nu_\mu$ events with high inelasticity, the generation for the training sample focused only on $\nu_\mu$ events. The training sample has a flat energy distribution from 1-200 GeV, in the target region for an oscillation analysis, and a nearly flat energy distribution from 200-500 GeV. CNNs are known for their difficulty extrapolating beyond their training samples \cite{NN-extrapolate}, so the energy is extended to 500 GeV to avoid issues at the sample boundary. To test the CNN, an independent GENIE sample was generated with $\sim4$~million unweighted $\nu_\mu$ CC events and $\sim1.6$~million unweighted $\nu_e$ CC events. These samples are weighted by both the atmospheric flux and standard oscillation parameters, so that the CNN is tested on data-like distributions.

\section{CNN Performance and Runtime}
\label{sec:results}

\begin{figure}[htbp]
\centering 
\includegraphics[width=.4\textwidth,scale=0.3]{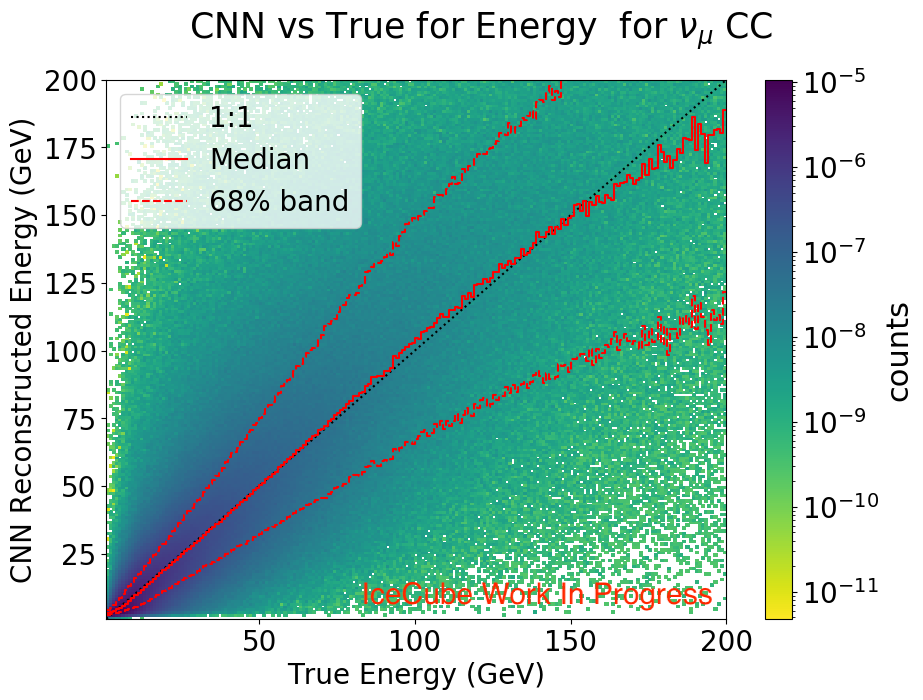}
\qquad
\includegraphics[width=.4\textwidth,scale=0.3]{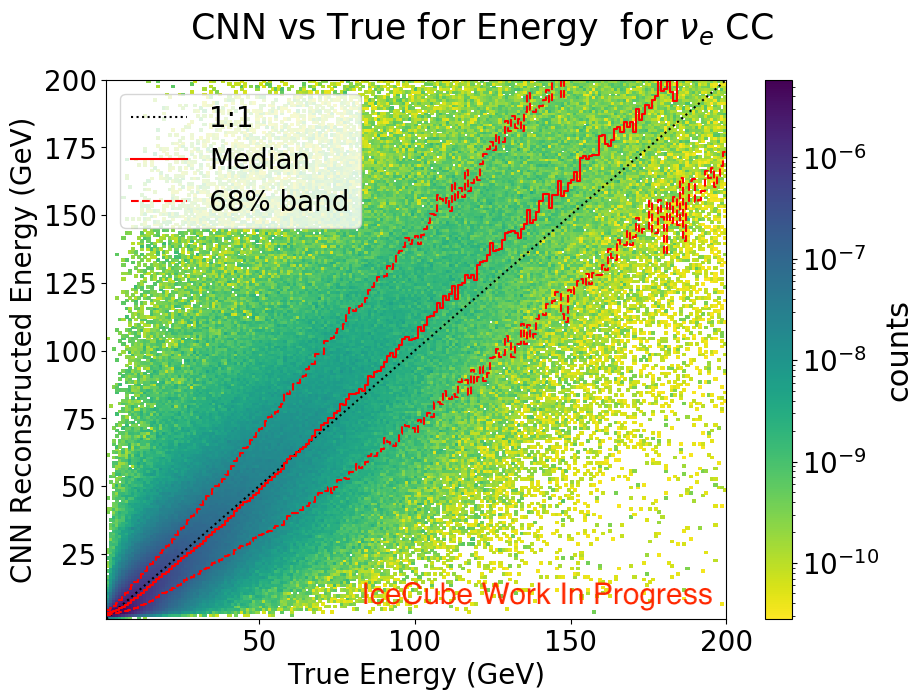}
\caption{\label{fig:Hist2D} CNN predicted energy vs. true energy for $\nu_\mu$ CC events (left) and $\nu_e$ CC events (right).}
\end{figure}

Figure \ref{fig:Hist2D} shows the CNN's energy prediction compared to the truth for the $\nu_\mu$ CC (left) and $\nu_e$ CC (right) testing samples. For the $\nu_\mu$ CC events, the CNN does well at low energies, with it's red, solid median line staying near the 1:1 perfect diagonal until about 150 GeV. At energies greater than 150 GeV, the CNN begins to slightly underestimate the true energy on average. At high energies, the performance is degraded by muon tracks that leave the volume of DOMs used for the CNN. Training the CNN with more high energy events generated above the target region (200-500 GeV) has improved resolution in past studies. Other options that are being explored for future improvements are expanding the CNN to include more IceCube DOMs or removing events that leave the detector volume from the final data set.

For the $\nu_e$ CC events, the CNN has a slight overestimation of the energy after 100 GeV. Since the CNN is trained on $\nu_\mu$ CC events only, it only sees the hadronic component of the cascade, which has less light per unit energy than electromagnetic cascades. This could account for the CNN's underestimation of the high energy $\nu_e$ events.
 
\begin{figure}[htbp]
\centering 
\includegraphics[width=.4\textwidth,scale=0.3]{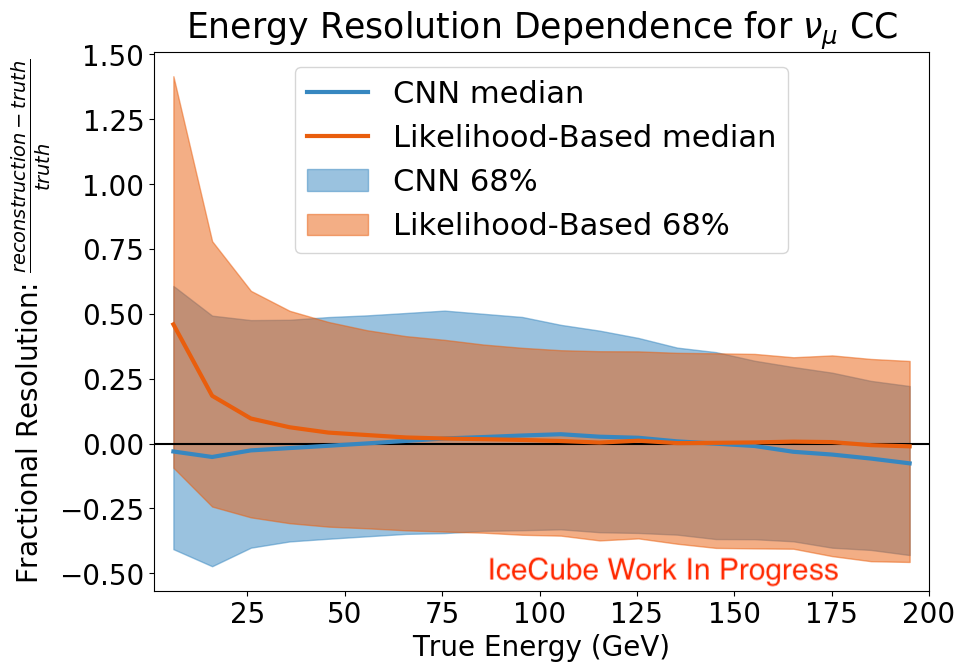}
\qquad
\includegraphics[width=.4\textwidth,scale=0.3]{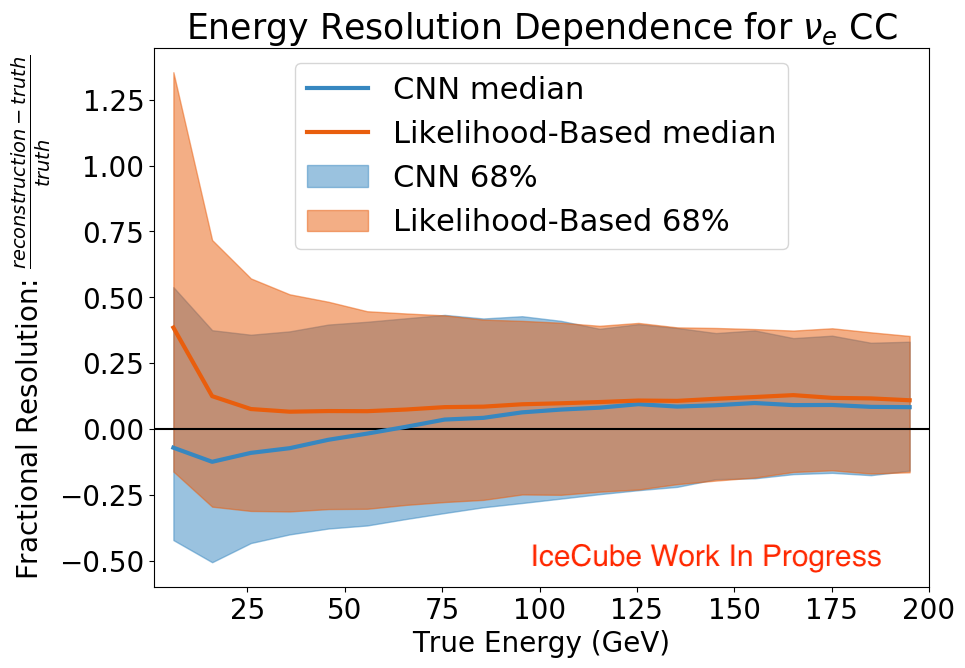}
\caption{\label{fig:Slices} Fractional energy resolution as a function of true neutrino energy, for $\nu_\mu$ (left) and $\nu_e$ (right).}
\end{figure}

Figure \ref{fig:Slices} shows a direct comparison of the energy-dependent fractional resolution for the CNN and the current likelihood-based reconstruction used for low-energy IceCube events. The CNN shows the most improvement at low energies (< 25 GeV) for both the $\nu_\mu$ CC (left) and $\nu_e$ CC (right) testing samples, where the majority of the atmospheric neutrino data is expected. The CNN also has comparable resolution with the likelihood-based method up to 200 GeV for both testing samples.

The CNN and likelihood-based reconstructions' runtimes are in Table \ref{tab:runtime}. Both the CNN's GPU and CPU runtimes are faster per event than the likelihood-based reconstruction (CPU capability only). It should be noted that the likelihood-based method reconstructs 8 variables simultaneously, but even if the CNN is run in serial to reconstruct 8 variables, its runtime would increase by less than a factor of 10. Speed is particularly important for IceCube's high statistics atmospheric data sample, which the CNN can reconstruct on a computer cluster in minutes.

\begin{table}[htbp]
\centering
\caption{\label{tab:runtime} Runtimes for the CNN (using GPU and CPU) and likelihood-based method (using CPU only)}
\smallskip
\begin{tabular}{|c | c | c | c |} 
 \hline
 \textbf{Method} & \textbf{Average Time per Event (seconds)} & \textbf{Events per Day per Single Core} \\[0.5ex] 
 \hline
CNN on GPU & 0.0077 & 11,000,000 \\
\hline
CNN on CPU & 0.27 & 320,000 \\
\hline
Likelihood-Based & 40  & 2,100 \\
\hline
\end{tabular} \\
\end{table}

\section{Conclusion}

The low-energy CNN has comparable energy resolution to current likelihood-based reconstruction methods and a speedup of $10^4$ in runtime when applied to 10-GeV scale IceCube events. The success of the low-energy CNN architecture extends beyond energy reconstruction, with ongoing explorations of applications spanning other variable reconstructions and classifications, such as neutrino direction \cite{Shiqi-proceedings}, which are necessary for neutrino oscillation analyses.

\acknowledgments

This material is based upon work supported by the National Science Foundation Graduate Research Fellowship Program under Grant No. DGE1848739 and additionally under NSF-1913607. Any opinions, findings, and conclusions or recommendations expressed in this material are those of the author(s) and do not necessarily reflect the views of the National Science Foundation.

\bibliographystyle{JHEP}
\bibliography{references}

\end{document}